\documentclass[reprint,amsmath,amssymb,floatfix,pra,aps,superscriptaddress]{revtex4-1}

\usepackage{graphicx}
\usepackage{amsmath}
\usepackage{amssymb}
\usepackage{upgreek}
\usepackage{subfigure}
\usepackage{bbold}
\usepackage[export]{adjustbox}
\usepackage[colorlinks=true,linkcolor = black,citecolor = blue]{hyperref}
\def\GHz{{\ \text{GHz}}}                     % GHz
\def\Rb87{^{87}\text{Rb}}                     % Rb 87
\def\Na23{^{23}\text{Na}}                     % Na 23
\def\Li6{^{6}\text{Li}}                       % Li 6
\def\bra#1{\mathinner{\langle{#1}|}}
\def\ket#1{\mathinner{|{#1}\rangle}}

\newcommand{\benum}{\begin{enumerate}}
\newcommand{\eenum}{\end{enumerate}}
\newcommand{\uw}{\rm MW}

\begin{document}

\title{Microwave Rabi resonances beyond the small-signal regime}
\author{A.~Tretiakov}
\affiliation{Department of Physics, University of Alberta, Edmonton AB, Canada}
\author{L.~J.~LeBlanc}
\affiliation{Department of Physics, University of Alberta, Edmonton AB, Canada}
\affiliation{Canadian Institute for Advanced Research, Toronto, ON, Canada}

\begin{abstract}
% Interactions between atoms and oscillating fields form the basis for many of the tools of atomic physics.
The coupling between microwave fields and atoms (or atom-like systems) is inherently weaker than for optical fields, making microwave signal manipulation for applications like quantum information processing technically challenging.  In order to better understand this coupling and to develop tools for measuring it, we explore the microwave coupling to atoms using the ``atomic candle'' technique, and push it beyond the bounds of the small-signal regime in order to deliver a larger signal.
In familiar two-level systems, responses beyond the usual Rabi oscillations can arise when  a single-tone drive is phase-modulated, causing the steady-state populations to oscillate at integer harmonics of the modulation frequency. Resonant behavior of the first two harmonics for frequencies near the Rabi frequency, known as $\alpha$ and $\beta$ Rabi resonances, is widely used for microwave-field magnetometry and  as a power standard known as the atomic candle.
Here, we explore Rabi resonances beyond the small-signal approximation and report upon experimental observations of higher-harmonic population response for microwave hyperfine transitions in cold $^{87}\rm Rb$ atoms, which we compare  to numerical simulations.  
\end{abstract}

\maketitle

\section{Introduction}

% Interactions between atomic vapors and microwave ($\uw$) fields play an important role in modern science and technology.
% Applications of these interactions are broad: in atomic clocks, these interactions underlie the current definition of the second~\cite{Bandi2012a,Pellaton2012,Petremand2012,Bandi2014}; they can be used for sensitive direct-current (DC)~\cite{Sun2018a} and  alternating current (AC)~\cite{Horsley2016} magnetometry;  and they are proposed as the mechanism for creating a transducer between microwave and optical frequencies for quantum information applications~\cite{Hafezi2012,Kiffner2016}. 

The control and manipulation of microwave signals with atoms is undergoing a renaissance in the scientific world, moving beyond traditional applications such as providing the definition of the second~\cite{Bandi2012a,Pellaton2012,Petremand2012,Bandi2014}, towards precision metrology in direct-current (DC)~\cite{Sun2018a} and  alternating current (AC)~\cite{Horsley2016} magnetometers, and novel applications for quantum information processing, such as microwave-to-optical transduction of qubits~\cite{Hafezi2012,Kiffner2016,Adwaith2019}. 
These two examples highlight two different ways atoms are used: first, as a measurement tool to detect the fields; and second, as the interaction medium through which the fields are manipulated.  Both cases demand an intimate understanding of the microwave-to-atom coupling, and a way to measure it precisely and with a large signal-to-noise ratio.

Very generally, an oscillating electromagnetic field with a frequency near the resonance of a two-level system periodically drives population transfer between  levels, in a process commonly known as Rabi oscillation.  In the presence of relaxation between these levels (due to one or more of many possible decoherence mechanisms), the steady-state populations tend towards a constant value.  In contrast, these steady-state populations oscillate when the phase of the oscillating field is modulated at a frequency $\omega_{\rm m}$, and  the amplitude of the population oscillations is enhanced when $\omega_{\rm m}$ is equal to the Rabi frequency, $\Omega_{\rm R}$.  This phenomenon, known as ``Rabi resonance''~\cite{Cappeller1985,Camparo1998,Camparo1998a}, can be used to find the  the system's Rabi frequency and, correspondingly, the power of the oscillating field at the position of the atom(s). 
In analogy to standard candles in astronomy, the Rabi-flopping atoms respond to AC and DC fields in exactly the same way in all laboratories, making them ``atomic candles''~\cite{Camparo1998a,Coffer2002}  for %measuring the absolute field strengths 
electromagnetic power standards.
% Hence, the Rabi frequency can be ''locked" to $\omega_m$ provided by a stable local oscillator, giving a MW~power standard, known as  ``atomic candle''~\cite{Camparo1998a,Coffer2002}. 
This   technique has also attracted significant interest for its applications for AC~magnetometry inside MW cavities~\cite{Sun2017a,Sun2018a}, MW wave guides~\cite{Kinoshita2009,Kinoshita2011,Kinoshita2013}, and in free space~\cite{Swan-Wood2001,Sun2018a,Liu2018,Shi2018,Kinoshita2017}. Applying static magnetic fields~\cite{Sun2017a} and using multispecies vapor cells~\cite{Sun2018a} adds frequency tunability and expands the operational bandwidth of this technique.   %and was recently used to obtain  resonance, reflection, and transmission characteristics of $\uw$ waveguide structures~\cite{Sun2017}. 

In previous work, the Rabi resonance was analyzed and measured only in the small-signal regime, where a sufficiently small modulation depth permits an approximate analytic solution to the Maxwell-Bloch equations for the populations.  These population oscillations, which have frequency components at  $\omega_{\rm m}$ and $2\omega_{\rm m}$, can be described analytically, including an amplitude that depends on $\omega_{\rm m}$. Here, we explore the atomic candle technique beyond the small-signal approximation both numerically and experimentally.  Using large-deviation phase modulation on the $\uw$ field, we observe higher-order spectral features whose characteristics and scaling behave differently than the first and second harmonics.  In this work, we focus our attention on the component oscillating at $4\omega_{\rm m}$, namely, the fourth harmonic, and discuss how it may be used to extend the atomic candle technique. 

% ****************************************************************************************************************************************************
%******* Model (Theoretical framework) *********
% ****************************************************************************************************************************************************

% ****************************************************************************************************************************************************
\section{Theory: Rabi oscillations with a phase-modulated driving field}

\begin{figure}[t!]  
\begin{center}
\includegraphics[]{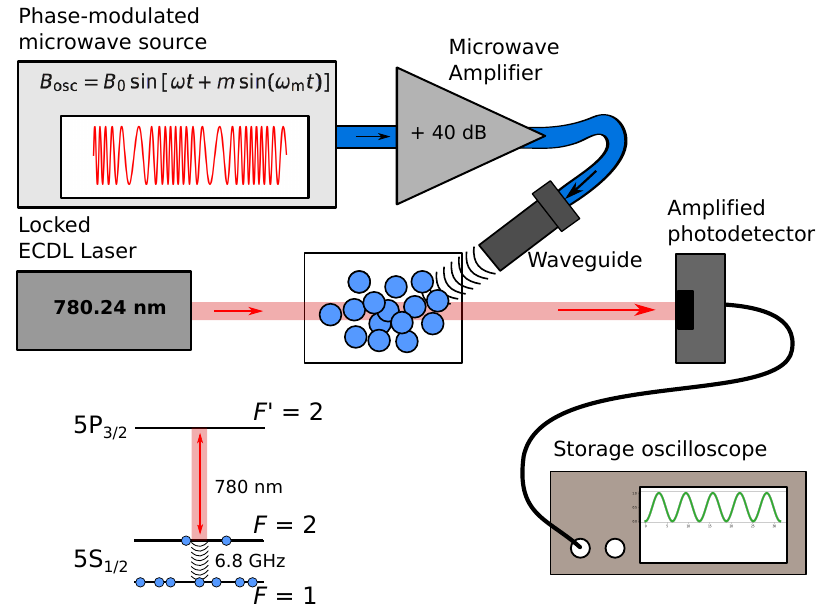} 
\caption{Schematic of phase-modulated microwave signals applied and measured in an atomic system, as used in experimental demonstrations. After optical molasses $^{87}$Rb atoms are optically pumped into $F=1$ state transparent for a probe light, resonant with $F=2\rightarrow F'=2$ transition. Change in the $F=2$ population  due to microwave field is monitored  in real time as a change in transmission signal of the probe light.}  
\label{fig:6}
\end{center}
\end{figure}

As we explore higher order Rabi resonances theoretically, we will consider the two-level system used in our experiments [Fig.~\ref{fig:6}], which is comprised of two Zeeman states in different hyperfine levels of an alkali metal, which are described by the total angular momentum $F$ and projections $m_F$.  For this  two-level system, separated by energy $\hbar \omega_0$, the levels can be coupled by a non-zero matrix element for magnetic dipole transitions, $V_{12} = \bra{2}\hat{\boldsymbol{\mu}}\cdot\mathbf{B}_{0}\ket{1}$, where $\hat{\boldsymbol{\mu}}$ is the magnetic dipole operator, and $\ket{1}$ and $\ket{2}$ represent the two levels.  Here, the AC driving field $\mathbf{B}(t)={B}_0\mathbf{e}_{B0}\cos\omega t$ is described by the peak field $\mathbf{B}_{0} = {B}_0\mathbf{e}_{B0}$, with AC-field amplitude ${B}_0$ and direction $\mathbf{e}_{B0}$.  When the AC field is resonant ($\omega = \omega_0$), the population oscillates at the Rabi frequency 
% $\Omega_{\rm R}= V_{12}/2\hbar$. 
$\Omega_{\rm R}= V_{12}/\hbar$.
When the detuning $\delta = \omega - \omega_0$ is non-zero, the oscillations occur at the generalized Rabi frequency is %$\Omega = \left(\Omega_{\rm R}^2 + \delta^2\right)^{1/2}$
$\Omega=\sqrt{\Omega_{\rm R}^2 + \delta^2}$.

%%%%%%%%%%%%%%%%%%%%%%%%%%%%%%%
%Figure 6
%%%%%%%%%%%%%%%%%%%%%%%%%%%%%%%%%%

In the presence of decoherence or damping, the oscillatory behavior of the two-level system damps out and the system equilibrates to a steady population ratio (e.g., equal populations for a strong resonant AC field and small damping). These dynamics are well-described by the Maxwell-Bloch equations~\cite{Noh2010}.

A phase-modulated oscillating field changes these dynamics.  Here, we consider the case where a time-dependent phase $\theta(t)$ modulates the AC field: $\mathbf{B}(t)={B}_0\mathbf{e}_{B0}\cos\left[\omega t + \theta(t)\right]$. In this case, the standard Maxwell-Bloch equations can be  modified to include additional time-dependent terms:
\begin{align}
\dot{u}&=[\delta+\dot{\theta}(t)]v-\Gamma_2u,\label{eq:OBE2u}\\
\dot{v}&=-[\delta+\dot{\theta}(t)]u+\Omega_{\rm R}w-\Gamma_2v,\label{eq:OBE2v}\\
\dot{w}&=-\Omega_{\rm R}v-\Gamma_1[w+1],
\label{eq:OBE2w}
\end{align} 
where $u$ and $v$ are the in- and out-of-phase coherences [$u =2 \rm Re(\rho_{12})$ and $v = 2\rm Im(\rho_{12})$] and $w = \rho_{22}-\rho_{11}$ is the population difference between levels, and where $\rho$ is the usual $2\times 2$ density matrix of the two-level system ($\rho_{11}$ is the ground-state population and $\rho_{22}$ is the excited-state population). In the loss terms, $\Gamma_1$ and $\Gamma_2$ represent longitudinal and transverse damping rates, respectively. We consider the case where the AC field is modulated about a constant phase offset $\theta_0$ with frequency $\omega_{\rm m}$, modulation depth $m$, and offset phase $\phi_{\rm m}$:
\begin{align}
\theta(t)=\theta_0+ m\sin(\omega_{\rm m}t+\phi_{\rm m}).
\label{eq:phase_modulation}
\end{align}

\begin{figure}[t!]  
\begin{center}
\includegraphics[]{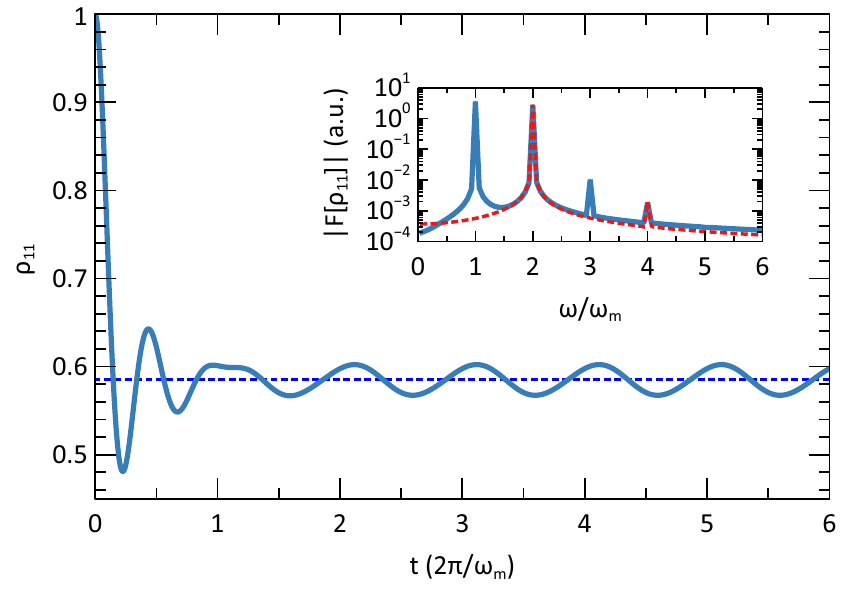} 
\caption{Ground state population  of a two-level system as a function of time in case of a phase-modulated coupling (solid line). The simulation is made for modulation depth $m=1$, 
% $\omega_{\rm m}=1$,  $\Omega_{\rm R}=2$, $\Gamma_1=0.8$, $\Gamma_2=0.4$, $\delta=1$. 
Rabi frequency $\Omega_{\rm R}=2\omega_{\rm m}$, longitudinal and transverse relaxation rates $\Gamma_1=0.8\omega_{\rm m}$ and $\Gamma_2=0.4\omega_{\rm m}$, respectively, carrier frequency detuning $\delta=\omega_{\rm m}$, where $\omega_{\rm m}$ is the modulation frequency.
Dashed curve corresponds to  steady-state solution without modulation. Inset: spectrum of the steady-state oscillations in the ground state population for $\delta=0.1\omega_{\rm m}$ (blue solid) and $\delta=0$ (red dashed), obtained using the Fast Fourier Transform (FFT).} 
\label{fig:1}
\end{center}
\end{figure}

We explore the numerical solutions of  Eqs.~\ref{eq:OBE2u}-\ref{eq:OBE2w} to find the time-dependent coherences and population dynamics under a variety of parameters.  In particular, we are interested in the  excited state population $\rho_{22} = (w+1)/2$ in the steady-state regime ($\Gamma_1t\gg1$), which is found to oscillate, even long after the damping times associated with $\Gamma_1$ and $\Gamma_2$ [FIG.~\ref{fig:1}].   

Below, we have a closer look at the time dependence of the steady-state populations of the ground and excited states, which is directly related to a measurable quantity: the absorption of light from one of these levels to an optically excitable third level.

% *********** Figure 1 % *********** 
% *********** end Figure 1 % *********** 

% ****************************************************************************************************************************************************
\subsection{Small-modulation approximation}
Treating the modulation term as a perturbation allows us to assume that in the steady-state regime the parameters describing the atomic state evolve around their unmodulated steady-state values $u_0,~v_0,~\rho_{22,0}$: 
\begin{align*}
    \rho_{22}(t)=\rho_{22,0}+\rho_{22,\rm m}(t),\\
    v(t) = v_0+v_{\rm m}(t),\\
    u(t) = u_0+u_{\rm m}(t),
\end{align*}
where $\rho_{22,\rm m}, v_{\rm m}, u_{\rm m}$ describe the evolution due to the phase modulation. 

FIG.~\ref{fig:1} shows that in the case of a weak modulation, the steady-state ground-state population $\rho_{11}(t)$ oscillates around its unmodulated value.   
Indeed, it can be shown~\cite{Coffer2002} that in this case the excited-state population evolution is described by
\begin{align*}
    \ddot{\rho}_{22,\rm m}+\Gamma_1\dot{\rho}_{22,\rm m}+&\Omega_{\rm R}^2\rho_{22,\rm m}=\\&2\Omega_{\rm R}\Gamma_2v_{\rm m}+2\Omega_{\rm R}(\delta+\dot{\theta})u_{\rm m}+u_0\Omega_{\rm R}\dot{\theta},
\end{align*}
which is a damped harmonic oscillator equation, where the driving is due to coherence between the levels ($v_{\rm m}$~and~$u_{\rm m}$) and phase modulation $\dot{\theta}$.  

For small modulations ($m<\sqrt{2\Gamma_1/\omega_{\rm m}}$ ) and low decoherence ($\Gamma_2\ll\omega_{\rm m}$), known as the small-signal approximation, the steady-state solution for the excited state population has been found analytically~\cite{Coffer2002}.  The population dynamics in this regime are represented by the first two harmonics of $\omega_{\rm m}$, and the excited-state dynamics with respect to its unperturbed value may be expressed as
\begin{align}
\rho_{22,m}(t)=P_1\sin(\omega_{\rm m}t+\phi_1)+P_2\sin(2\omega_{\rm m}t+\phi_2),
\label{eq:alpha_resonance}
\end{align}
with amplitudes 
\begin{align}
P_1&=\frac{\tfrac{m}{2}\omega_{\rm m}\Omega_{\rm R}^2\delta}{\left[\Gamma_2^2+\delta^2+\frac{\Gamma_2}{\Gamma_1}\Omega_{\rm R}^2\right]\sqrt{\left(\omega_{\rm m}^2-\Omega_{\rm R}^2\right)^2+\Gamma_1^2\omega_{\rm m}^2}},\\
P_2&=\frac{\left(\tfrac{m}{2}\right)^2\omega_{\rm m}\Omega_{\rm R}^2\Gamma_2}{\left[\Gamma_2^2+\delta^2+\frac{\Gamma_2}{\Gamma_1}\Omega_{\rm R}^2\right]\sqrt{\left(\omega_{\rm m}^2-4\Omega_{\rm R}^2\right)^2+4\Gamma_1^2\omega_{\rm m}^2}}.
% \\
% P_1\sim \delta
\label{eq:beta_resonance}
\end{align}
The main feature of these solutions is that the amplitudes $P_1(\omega_{\rm m},\Omega_{\rm R})$ and $P_2(\omega_{\rm m},\Omega_{\rm R})$ peak when the frequency of the corresponding harmonic is resonant with the Rabi frequency, $\omega_{\rm m} = \Omega_{\rm R}$ and $2\omega_{\rm m} = \Omega_{\rm R}$, respectively, which are known as the $\alpha$ and $\beta$ Rabi resonances.   Experimentally, scanning the modulation frequency $\omega_{\rm m}$ in search of this peak provides a tool for measuring $\Omega_{\rm R}$ and, thus, the power of the driving field, since $\Omega_{\rm R} \propto B_0^2$~\cite{Sun2018a}.  In addition, the amplitude of the first harmonic disappears at zero-carrier-detuning, $\delta = 0$ ($P_1\propto \delta$).  By tuning the carrier frequency $\omega$ to field-sensitive transitions~\cite{Sun2018a}, this dependence provides a means by which to measure static magnetic fields.

% #################################################

% Figure 2
\begin{figure}[t!]  
\begin{center}
\includegraphics[]{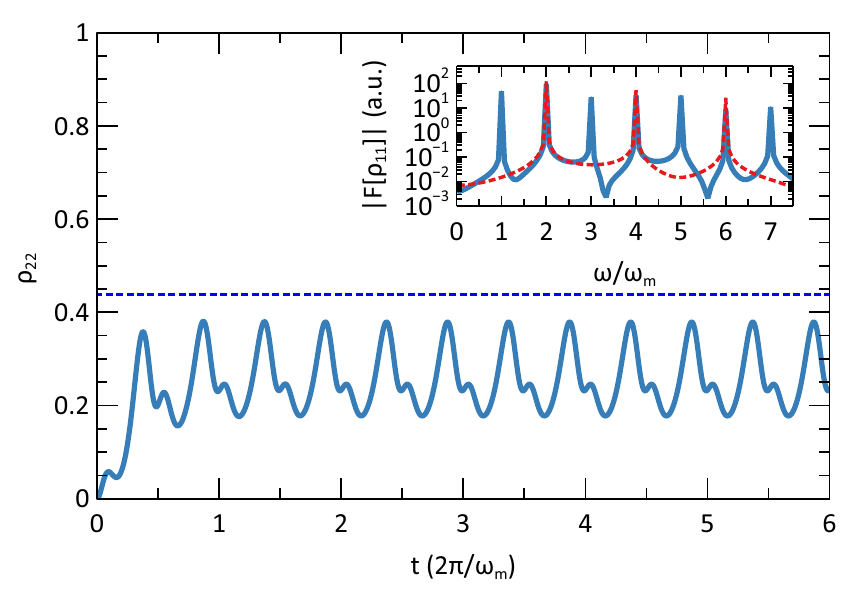}
\caption{Excited state population  of a two-level system as a function of time in case of a phase-modulated coupling (solid line) with $m=6$~rad, 
% $\omega_{\rm m}=1$,  $\Omega_{\rm R}=1.5$, $\Gamma_1=0.8$, $\Gamma_2=0.4$, $\delta=0$. 
$\Omega_{\rm R}=1.5\omega_{\rm m}$, $\Gamma_1=0.8\omega_{\rm m}$, $\Gamma_2=0.4\omega_{\rm m}$, $\delta=0$. 
Dashed line corresponds to  steady-state solution without modulation. Inset: spectrum of the steady-state oscillations in the excited state population for $\delta=\omega_{\rm m}$ (blue solid) and $\delta=0$ (red dashed).} 
\label{fig:oscillations_simulation_large}
\end{center}
\end{figure}

% End Figure 2
%#####################################################

The inset of FIG.~\ref{fig:1} compares the spectra of the  steady-state $\rho_{11}$, simulated for $\delta=0$ and $\delta=0.1\omega_m$. As is expected, the first harmonic vanishes at zero-carrier-detuning. In the simulation, the magnitudes of the decoherence rates are comparable with the modulation frequency, which violates the small-signal approximation and leads to additional oscillations at third and fourth harmonics, even though their amplitudes are a few orders of magnitude smaller than~$P_1$~and~$P_2$.

% #################################################

% Figure 3
%##########################################################
\begin{figure}[t!]  
\begin{center}
\includegraphics[]{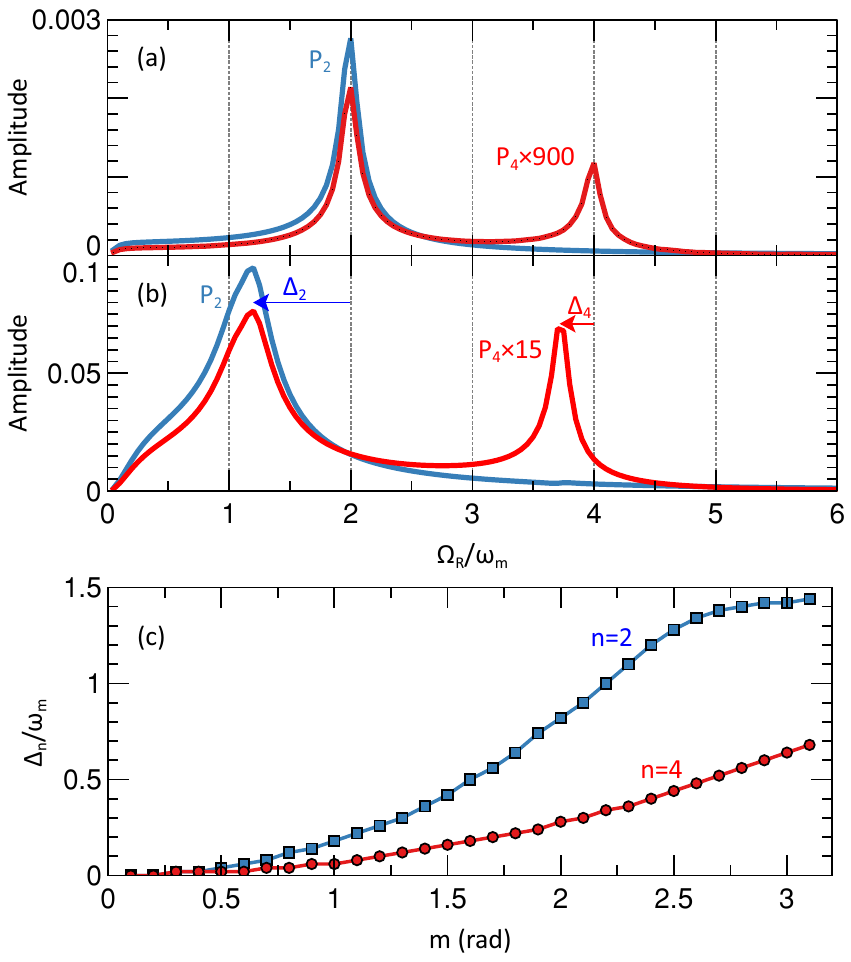}
\caption{(a) and (b): Rabi resonances  for amplitudes of  $2^{\rm nd}$ (P$_{2}$)  and $4^{\rm th}$ (P$_{4}$)  harmonics for $m=0.25$~rad and $m=2$~rad, respectively. The amplitude of the $4^{\rm th}$ harmonic is scaled for better visibility, as indicated. At higher $m$ peaks shift with respect to their original positions $\Omega_{R}=2\omega_{\rm}$ and $\Omega_{R}=4\omega_{\rm}$, which indicated by $\Delta_2$ and $\Delta_4$, respectively. (c) Shifts of the Rabi resonance peaks with respect to their weak-modulation positions as a function of the modulation index. Simulation is made for $\Gamma_1=2\Gamma_2=0.08\omega_{\rm m}$ and $\delta=0$. } 
\label{fig:3}
\end{center}
\end{figure}

%End Figure 3
%#############################################

%%%%%%%%%%%%%%%%%%%%%%%%%%%%%%%%%%%%%%%%%%%%%%%
%%%%%%%%%%%%%%%%%%%%%%%%%%%%%%%%%%%%%%%%%%%%%%%
%%%%%%%%%%%%%%%%%%%%%%%%%%%%%%%%%%%%%%%%%%%%%%%
\subsection{Large modulation: Numerical analysis}
\label{sec:large modulation}

A general solution of the phase-modulated oscillatory dynamics of a two-level system, with arbitrary phase- and amplitude-modulation, was obtained in Ref.~\cite{Alekseev1992}.

Still, there is no simple analytical expression for the case where the  phase is modulated according to Eq.~\ref{eq:phase_modulation}.  
Numerical simulations of the large-modulation condition  [FIG.~\ref{fig:oscillations_simulation_large}] show that the steady-state ground-state population has a periodic solution, but, unlike in the small-modulation case, its average is shifted from the unmodulated value and its spectrum  consists of many harmonics at multiples of $\omega_{\rm m}$:
\begin{align*}
    \rho_{11}(t)=\sum_nP_n\sin(n\omega_mt+\phi_n),
\end{align*}
where $n$ can be any positive integer and $\phi_n$ is the phase of the response.
As might be expected from small-signal theory, the relative height of the spectral consituents $P_n$ depends on $m, \omega_{\rm m}, \Omega_{\rm R}, \Gamma_{1,2}$ and $\delta$. In the case when $\delta=0$, all \emph{odd} components disappear, which is consistent with the analytical small signal solutions (Eq.~\ref{eq:alpha_resonance}).  However, compared to the small signal solution, the large-modulation dynamics include additional spectral peaks, and thus more (correlated) information from which to extract the field calibration, %making the final result more precise
providing more options for extracting the final results.

Next, we evaluate the response of the  excited-state population when the carrier frequency is equal to the frequency of the transition between the ground and excited states (i.e., $\delta = 0$), which is the regime of practical interest.  Here, the ground state population's response to phase modulation includes only the even harmonics of $\omega_{\rm m}$ and we study, in particular, the second and fourth harmonics for different values of the modulation index~$m$. We are interested only in the steady-state response, where $\Gamma_1t_1\gg 1$ and any ``normal'' Rabi oscillations would have damped out. 

To extract values of the spectral amplitudes $P_n$ of the steady-state ground-state population signal $\rho_{11}(t)$ from our numerical simulations, we consider both the in- and out-of-phase responses of the signal at frequency $n\omega_{\rm m}$: 
\begin{align*}
    a_n=\dfrac{1}{N}\sum_{k=1}^N\rho_{11}(t_k)\sin n\omega_{\rm m}t_k,\\
    b_n=\dfrac{1}{N}\sum_{k=1}^N\rho_{11}(t_k)\cos n\omega_{\rm m}t_k,
\end{align*}
where  %we use a Fourier-series approach to determine the spectral components such that
the $t_k$ represent the evenly-spaced discrete time points, and the index $k$ runs over all time indices. This  %Fourier-series
approach to the analysis provides the amplitudes of the components at each harmonic of the modulation frequency $\omega_{\rm m}$, and the overall overall amplitude is given by $P_n=2\sqrt{a_n^2+b_n^2}$.

This analysis relies on the fact that $a_n$~and~$b_n$ approach zero as $N$ approaches infinity for any frequency  other than the reference frequency (i.e., $\omega\neq n\omega_{\rm m}$). In the case when $N$ is finite, the contribution from these unwanted frequencies can be significant, so we filter them out in the frequency  domain  using  Fast  Fourier  transform  and  its  inverse. To reduce the contribution of the numerical artifacts due to finite duration of the analyzed signal, we apply a Hamming window to $\rho_{11}(t)$ before calculating $a_n$~and~$b_n$.

%########################################
\subsubsection{Rabi resonance: low decoherence rate}
%########################################

Figure~\ref{fig:3}(a) shows the simulated amplitude response for the second and fourth harmonics as a function of the Rabi frequency with fixed $\omega_{\rm m}$ and $\Gamma_1, \Gamma_2\ll \omega_{\rm m}$. In agreement with Eq.~\ref{eq:beta_resonance}, the amplitude $P_2$ of the second harmonic peaks when $\Omega_{\rm R}=2\omega_{\rm m}$.
The fourth harmonic has two  resonances:  at  $\Omega_{\rm R} = 2\omega_{\rm m}$ and $\Omega_{\rm R} = 4\omega_{\rm m}$. The latter peak is  similar to the frequency response of a damped harmonic oscillator whose natural frequency is $\Omega_{\rm R}$ and is driven harmonically at $4\omega_{\rm m}$. In contrast, the resonant behavior of the second and fourth harmonics in the case of large phase modulation  deviates from this small signal approximation, as seen in FIG.~\ref{fig:3}(b). Both harmonics' resonance peaks shift toward smaller values of $\Omega_{\rm R}/\omega_{\rm m}$, but the shift  of $P_2$'s peak (i.e., $\Delta_2$) is much more significant than that of $P_4$ (i.e., $\Delta_4$).  Additionally,  its shape changes drastically, including an increase in the linewidth. The value of this shift as a function of $m$ is shown in FIG.~\ref{fig:3}(c), which indicates that the fourth-harmonic Rabi resonance is more stable against variation of the modulation depth.

\begin{figure}[t!]  
\begin{center}
\includegraphics[]{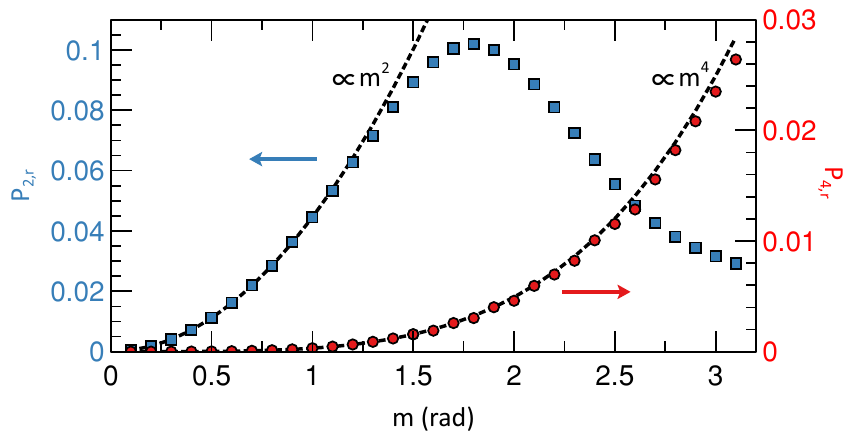}
\caption{Height of the Rabi resonance peaks of $2^{\rm nd}$ (blue squares) and $4^{\rm th}$ (red circles) as a function of modulation depth, determined from numerical simulation with $\Gamma_1=2\Gamma_2=0.08\omega_{\rm m}$ and $\delta=0$. The value for $4^{\rm th}$ harmonic ($P_{4,\rm r}$) is calculated for the right-hand peak on FIG~\ref{fig:3}(a-b), which corresponds to $\Omega_{\rm R}\approx4\omega_{\rm m}.$ Black dashed curves correspond to a quadratic and a quartic functions normalized to $P_{2,\rm r}(m=1)$ and $P_{4,\rm r}(m=1)$, respectively, to indicated the corresponding power dependence $P_{2,\rm r}$ and $P_{2,\rm r}$ at lower $m$. }  
\label{fig:4}
\end{center}
\end{figure}

Furthermore, the height $P_{4,\rm r}$ of the fourth-harmonic Rabi-resonance peak corresponding to $\Omega_{R}=4\omega_{\rm m}$ scales as $m^4$,  compared to quadratic ($m^2$) dependence of the second harmonic's peak value $P_{2,\rm r}$ [FIG.~\ref{fig:4}] (note that the quadratic behavior also fails at large $m$, indicating that Eq.~\ref{eq:beta_resonance} is no longer valid). Therefore, in the large-phase-modulation regime, $P_4$'s peak is comparable to $P_2$'s, and thus it becomes advantageous to use a fourth-harmonic atomic candle (4HAC), where the Rabi resonance of the fourth harmonic is used.%, by which to stabilize~$\Omega_{\rm R}$.

%%%%%%%%%%%%%%%%%%%%%%%%%%%%%%%
%Figure 4
%%%%%%%%%%%%%%%%%%%%%%%%%%%%%%%%%%

%%%%%%%%%%%%%%%%%%%%%%%%%%%%%%%
% end Figure 4
%%%%%%%%%%%%%%%%%%%%%%%%%%%%%%%%%%

%########################################
\subsubsection{Rabi resonance: high decoherence rate}
\begin{figure}[h!]  
\begin{center}
\includegraphics[]{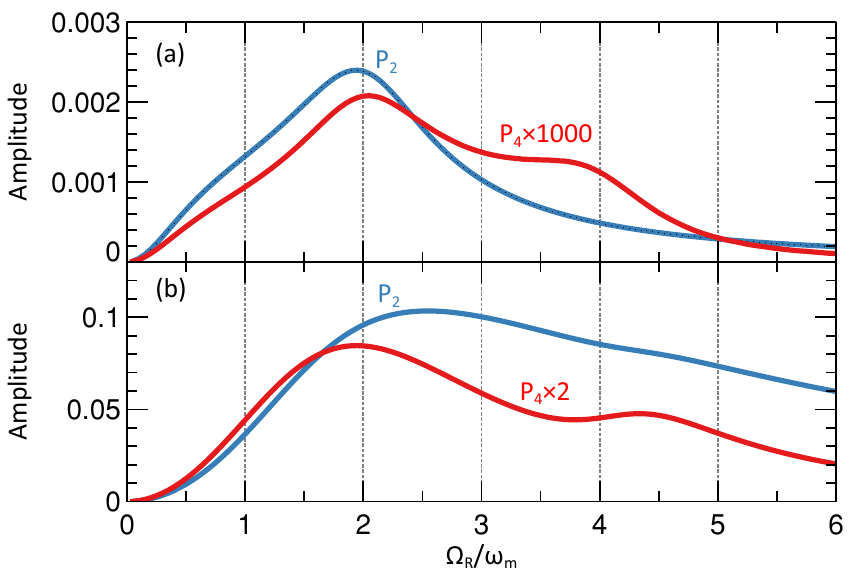}
\caption{Rabi resonances for $2^{\rm d}$  and $4^{\rm th}$ harmonics in the case large decoherence. (a)~and~(b) correspond to  $m=0.25$~rad and $m=6$~rad, respectively. Simulation is made for $\Gamma_1=2\Gamma_2=0.8\omega_{\rm m}$ and $\delta=0$. The amplitude of the $4^{\rm th}$ harmonic is scaled for better visibility.} 
\label{fig:5}
\end{center}
\end{figure}

%########################################
According to Eq.~\ref{eq:beta_resonance}, the width of $P_2$-peak should increase with increasing decoherence rates $\Gamma_{1,2}$. FIG.~\ref{fig:5} shows that this applies to both $P_4$-peaks as well, and when the decoherence rate is comparable to  $\omega_{\rm m}$, these two peaks overlap. Even at small $m$, the $P_2$ peak's shape is different from  a Lorentzian (as in  FIG.~\ref{fig:3}), which indicates that Eq.~\ref{eq:beta_resonance} is no longer valid. For stronger modulation, line-broadening and asymmetry make both features unviable for practical applications.   
%%%%%%%%%%%%%%%%%%%%%%%%%%%%%%%
%Figure 5
%%%%%%%%%%%%%%%%%%%%%%%%%%%%%%%%%%

%%%%%%%%%%%%%%%%%%%%%%%%%%%%%%%%%%
%%%%%%%%%%%%%%%%%%%%%%%%%%%%%%%%%%
%%%%%%%%%%%%%%%%%%%%%%%%%%%%%%%%%%%%
\section{Experiment}
\subsection{Experimental setup}

% %%%%%%%%%%%%%%%%%%%%%%%%%%%%%%%
% %Figure 6
% %%%%%%%%%%%%%%%%%%%%%%%%%%%%%%%%%%

% \begin{figure}[t!]  
% \begin{center}
% \includegraphics[]{FIG6.pdf} 
% \caption{Setup concept. After optical molasses $^{87}$Rb atoms are optically pumped into $F=1$ state transparent for a probe light, resonant with $F=2\rightarrow F'=2$ transition. Change in the $F=2$ population  due to $\uw$ field is monitored  in real time as a change in transmission signal of the probe light.}  
% \label{fig:6}
% \end{center}
% \end{figure}

%%%%%%%%%%%%%%%%%%%%%%%%%%%%%%%
% end Figure 6
%%%%%%%%%%%%%%%%%%%%%%%%%%%%%%%%%%

To verify the findings of our numerical calculations, we tested the fourth-harmonic atomic candle~(4HAC) technique using our cold-atom apparatus~[FIG.~\ref{fig:6}].  

In our experiment we use a cloud of laser-cooled $^{87}$Rb atoms, which have undergone  standard magneto-optical trapping, followed by an ``optical molasses'' step giving a 1-mm-wide cloud with typical atom number of $10^8$ and temperature of $70\ \mu$K.  The atoms are optically pumped into $F=1$ ground state by switching off the repumping light 1~ms before the cooling and trapping beams are turned off and the atoms begin to expand in time-of-flight. Next, the cloud is illuminated by a travelling $\uw$ field  emitted from an open-ended rectangular waveguide, whose phase is modulated periodically with frequency $\omega_{\rm m}$  using built-in functions of the $\uw$ source (SRS SG384). The near-resonant $\uw$ field (with carrier frequency $\omega/2\pi \approx 6.8~\GHz$) transfers the population between the ground states $F=1$ and $F=2$.  This transfer is observed by monitoring the absorption of of a weak probe laser beam resonant with $F=2\rightarrow F'=2$ transition~[FIG.~\ref{fig:6}], which is proportional to the population of state $F=2$. Since the best signal is produced when both the optical and MW~fields are on resonance with the corresponding transitions, this technique is known as the double resonance~(DR)~\cite{Bandi2012a}. The transmission signal is measured by a high-sensitivity photodetector and recorded with an oscilloscope.  During the laser trapping and cooling stages, the $\uw$ source is detuned by $\delta/2\pi =  100$~MHz from the hyperfine splitting, but during the DR~interrogation it is quickly brought back to resonance by mixing with an external 100~MHz radio-frequency signal. The $\uw$ signal is amplified by a separate amplifier, providing about $2$~W of power to the waveguide.

To measure the response of the atoms to the modulated microwave fields, and to test the 4HAC technique, we systematically varied  the carrier detuning $\delta$ to find the resonant condition.  Next, once resonance was found ($\delta = 0$), the phase-modulation frequency was varied to find the Rabi-resonance condition  $\delta_{\rm m} = 0$.  The time-dependent transmission of the resonant optical beam served as a measure of the population in the $F = 1$ ground state, and a frequency analysis of the transmission dynamics was performed numerically following the data acquisition.

%%%%%%%%%%%%%%%%%%%%%%%%%%%%%%%
%Figure 7
%%%%%%%%%%%%%%%%%%%%%%%%%%%%%%%%%%

\begin{figure*}[tb!]  
\includegraphics[]{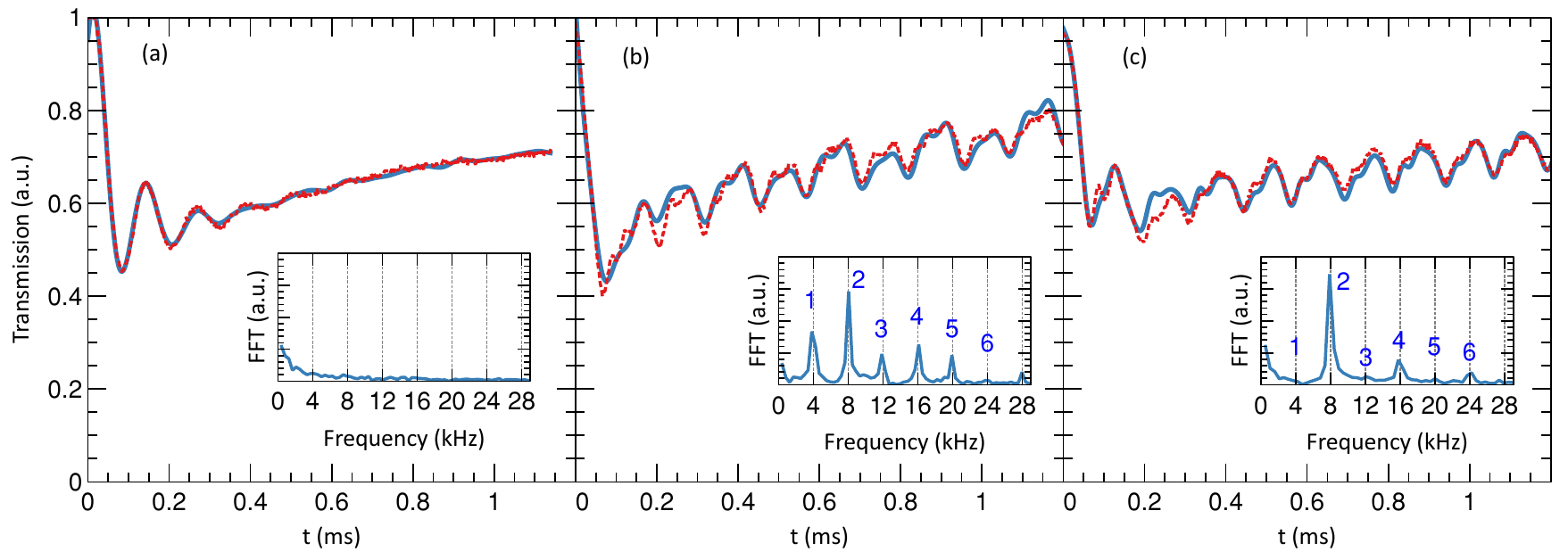} 
\caption{Typical DR~signal for Rabi oscillations as a function of time  with a fit according to the model described in the text (red dashed line) and an FFT~spectrum of the tail of the signal (inset) for various parameters of the driving $\uw$ field. (a)~Unmodulated oscillations; 
% (b)~$\omega_{\rm m} = 4$~kHz, $m = 2\pi$, $\delta = 7$~kHz;   
(b)~$\omega_{\rm m}/2\pi = 4$~kHz, $m = 2\pi$, $\delta/2\pi = 7$~kHz; 
% (c)~$\omega_m = 4$~kHz, $m = 2\pi$, $\delta\approx0$. 
(c)~$\omega_{\rm m}/2\pi = 4$~kHz, $m = 2\pi$, $\delta\approx0$. 
Axes in all insets have the same scaling. (b) shows that, in general,  in case of a phase-modulated driving field the steady-state contains  harmonics of the modulation frequency (4~kHz), compared to a flat spectrum in case of an unmodulated driving field. From (c), it is seen that the odd harmonics disappear when the carrier frequency of the driving field is resonant with the transition. The integer numbers next to peaks in the insets indicate harmonics of $\omega_{\rm m}$ the peaks correspond to. For clarity,  only first $1.5$~ms of the DR is signal is shown. The probe's power is 46~$\mu$W.} 
\label{fig:7}
\end{figure*}

%%%%%%%%%%%%%%%%%%%%%%%%%%%%%%%
%End Figure 7
%%%%%%%%%%%%%%%%%%%%%%%%%%%%%%%%%%
%%%%%%%%%%%%%%%%%%%%%%%%%%%%%%%%%%%%%%%%
%%%%%%%%%%%%%%%%%%%%%%%%%%%%%%%%%%%%%%%%
\subsection{Results and analysis}
\label{sec:Results and analysis}
Using large modulation depths, phase-modulated MW signals applied to our cold atomic samples revealed multi-harmonic nature of the steady oscillations beyond the small-signal regime.
FIG.~\ref{fig:7} shows the first $1.2$~ms of the measured transmission signals (total duration is $5$~ms) and derived spectra for three different conditions in our atomic candle experiments. First, we demonstrate ``pure'' Rabi oscillations [FIG.~\ref{fig:7}(a)] by applying a resonant but unmodulated $\uw$ field.  The spectrum here is featureless.  Next, when  $\uw$-field is strongly modulated but detuned from the hyperfine transition ($m=2\pi$, $\delta \neq 0$), the population and thus the transmission signal have additional modulations [FIG.~\ref{fig:7}(b)], and the spectrum contains peaks at integer values of $\omega_{\rm m}$.  Finally, when the carrier frequency is resonant ($\delta = 0$),  the odd harmonics of the frequency of the phase modulation  are suppressed in the transmission signal [FIG.~\ref{fig:7}(c)]. We observed that the amplitudes of the spectral components  slightly varied from shot to shot under constant parameters of the \uw~field setup, and during some of the shots we do not see the total elimination of the odd harmonics. This might be due to noise in the response of the radio-frequency mixer, and for later analysis of the amplitudes we take average values over a few trials.

%%%%%%%%%%%%%%%%%%%%%%%%%%%%%%%
%Figure 8
%%%%%%%%%%%%%%%%%%%%%%%%%%%%%%%%%%

\begin{figure}[b!]  
\includegraphics[]{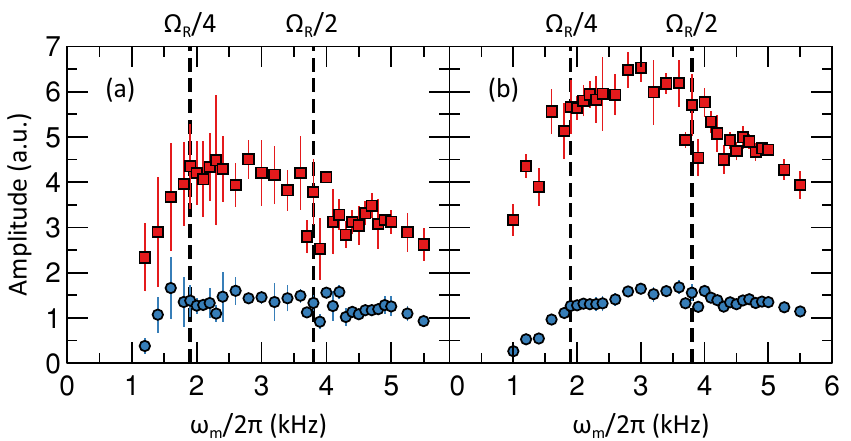}
\caption{Rabi resonance features of the second (red squares) and fourth (blues circles) harmonics for $m=2\pi, \delta=0$ and the probe's power of 46~$\mu$W. Each point and its error-bar correspond to an average and variance of $5$~values extracted from a tail of a DR~signal, respectively. (a)~Amplitudes were found by using a quadrature detection discussed in Section~\ref{sec:large modulation}. (b)~Amplitudes were found by fitting the DR~signal to a model described in Section~\ref{sec:Results and analysis}. Two vertical dashed lines correspond~to~$\omega_{m}=\Omega_{\rm R}/4$ and~$\omega_{m}=\Omega_{\rm R}/2$.} 
\label{fig:8}
\end{figure}
%%%%%%%%%%%%%%%%%%%%%%%%%%%%%%%
%End Figure 8
%%%%%%%%%%%%%%%%%%%%%%%%%%%%%%%%%%
Because our experiment allows us to record only a very limited number of oscillations in the time-domain, spectral analysis using numerical 
FFTs does not yield a consistent set of results. Instead, we retrieve the spectral information by using the same numerical routine as in~Section~\ref{sec:large modulation}. For additional validation we fit the time-series of the transmission signal $f(t)$ to a model 
\begin{align*}
f(t)&=\sqrt{A^2+(Bt)^2}+C_1e^{-\Gamma_1t}\\
&+C_2e^{-\Gamma_2t}\sin^2\left(\dfrac{\Omega_{R\rm }}{2}t+\phi\right)
+\sum_{n=1}^{6}P_n\sin(n\omega_mt+\phi_n),
\label{eq:fit}
\end{align*}
where the first term accounts for the loss of atoms due to interaction with the probe; the  second term accounts for  longitudinal damping; the third describes damped Rabi oscillations, and the final terms correspond population dynamics at the harmonics of the modulation frequency.  In this model,  $A,~B,~C_{1,2},~P_n,~\phi,~\phi_n$ are all fit parameters, where: $A$ represents the initial density of atoms at the position of the probe; $B$ is the rate of hole burning due to momentum kicks from the probe; $C_1$ and $C_2$ the amplitude of the transmission  oscillations, which are proportional to the amplitude oscillations; $\phi$ and $\phi_{\rm n}$ represent the initial phase of the carrier and modulation frequencies at the start of the transmission measurement; and $P_n$ are the steady-state amplitudes of the population oscillation components that we are looking for in this atomic candle measurement.

To investigate practicability of the 4HAC, we analyzed the amplitudes of the second and fourth harmonics, $P_2$ and $P_4$ as a function of the modulation frequency $\omega_{\rm m}$ at a fixed $\uw$ power.  (For these analyses,  we use only the tail of the data, which corresponds to the steady-state regime.)  FIG.~\ref{fig:8} shows that the amplitudes of both harmonics  peak as $\omega_{\rm m}$ is scanned through $\Omega_{\rm R}/2 = 2\pi\times(3.8\pm0.3)$~kHz in a manner similar to FIG.~\ref{fig:5}(b), where $\Omega_{\rm R}$ was estimated from an analysis of a set of unmodulated signals. The uncertainty in $\Omega_{\rm R}$ is largely of the same origin as the uncertainties in the amplitudes (as was discussed before), since when the carrier-wave is not exactly on resonance with the transition, the oscillation response is at a higher frequency. By fitting a set of transmission data corresponding to unmodulated  driving field, we estimate the decoherence rates in our experiment to be
% $\Gamma_1=3.2\pm0.2$~kHz and $\Gamma_2=1.55\pm0.07$~kHz. 
$\Gamma_1=(2.0\pm0.1)\times10^4$~s$^{-1}$ and $\Gamma_2=(0.97\pm0.04)\times10^4$~s$^{-1}$. 
Even though in this regime  Rabi resonances are not practically useful, the qualitative behavior supports our model in numerical analysis.

%%%%%%%%%%%%%%%%%%%%%%%%%%%%%%%%%%%%%%%%
%%%%%%%%%%%%%%%%%%%%%%%%%%%%%%%%%%%%%%%%
\subsection{Experimental considerations}

In~\cite{Coffer2002} it was shown that increasing the probe's power leads to a broadening of the line shapes of the atomic-candle Rabi resonances. The absorption and  subsequent spontaneous emission of a photon by the atom from the upper ground state contributes to the  longitudinal relaxation, and thus $\Gamma_1$ is proportional to the absorption rate, which in turn is proportional to the probe's intensity.

\begin{figure*}[h]  
\includegraphics[]{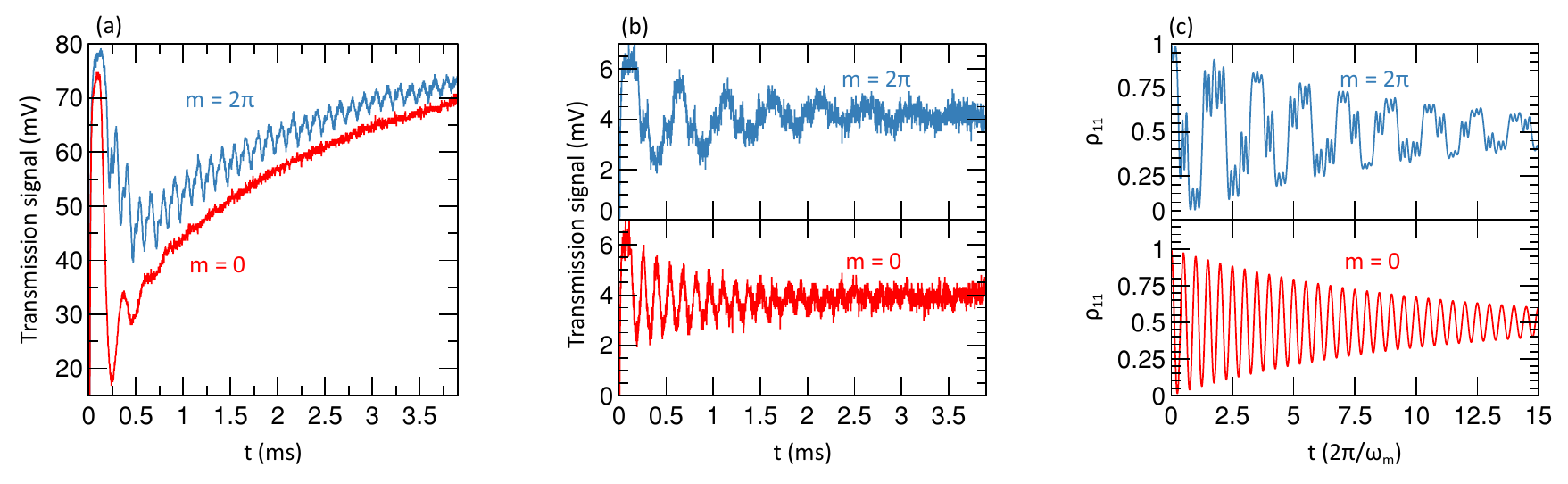}
\caption{(a-b): Influence of the probe's power on the double resonance signal in case of a phase-modulated (blue) and unmodulated (red) driving $\uw$~field. (a) Probe's power is $0.54\ \mu$W; (b) Probe's power is $0.06\ \mu$W; (c):  numerical simulation of a two-level system evolution in the case of  small damping corresponding to (b), with simulation parameters 
% $\omega_{\rm m}=1,$ $\Omega_{\rm R}=2$, $\Gamma_1=2\Gamma_2=0.02$.}
$\Omega_{\rm R}=2\omega_{\rm m}$, $\Gamma_1=2\Gamma_2=0.02\omega_{\rm m}$.}
\label{fig:9}
\end{figure*}

In our experiment we can observe a similar effect, where by varying the power of the probe beam  we can obtain qualitatively different oscillation patterns. FIG.~\ref{fig:9}~shows that increased probe's power leads to a higher decoherence rate and changing it by a factor of $10$ from $0.54\ \mu$W to  $0.06\ \mu$W brings the system from highly-damped~[FIG.~\ref{fig:9}(a)] to underdamped~[FIG.~\ref{fig:9}(b)] oscillations. This is in a good qualitative agreement with our simulations, where the damped case corresponds to FIG.~\ref{fig:oscillations_simulation_large}, and a simulation of the underdamped case is shown~in~[FIG.~\ref{fig:9}(c)]. Since our simulation model is based on  the atom as a two-level system and does not describe the absorption of the probe, it does not capture the monotonically increasing transmission over time that is experimentally observed in the case of the higher probe power. The fact that the average transmission level does not change significantly in the case of  the smaller probe intensity indicates that the nonzero steady-state slope of the transmission signal is due to the hole-burning in the cloud rather than thermal expansion.  
%%%%%%%%%%%%%%%%%%%%%%%%%%%%%%%%%%%

For our steady-state analysis, we want to work in a regime where the Rabi oscillations are damped. 
Since in our case the interrogation time is limited, this can be achieved by using a higher probe power, corresponding to  a strongly-damped regime. In addition, higher probe power gives a better signal~to~noise~ratio.

The main limitation in our experiment was the short interrogation time, limited by thermal expansion of the atomic cloud and interaction with the probe beam, which was ``burning'' a hole in the cloud. In addition, we observed slight discrepancies between several successive measurement of the DR~signal at constant $\uw$~field parameters, which are probably due to fluctuations in the turn-on time of the field. These should not be an issue in the real-world configurations, such as vapor cells, where interaction with the buffer gas keeps the atoms in place, or in cold atoms whose interrogation time is increased by adding a trapping potential, e.g., an optical dipole trap. Another issue is that, due to the geometry of our experiment, it is hard to select a specific Zeeman transition by applying a bias magnetic field with a particular direction with respect to the polarization of the $\uw$ field. Again, in vapor-cell applications this problem usually is not present. 

%%%%%%%%%%%%%%%%%%%%%%%%%%%%%%%
%Figure 9
%%%%%%%%%%%%%%%%%%%%%%%%%%%%%%%%%%

%%%%%%%%%%%%%%%%%%%%%%%%%%%%%%%%%%%%%%%% 
%End Figure 9
%%%%%%%%%%%%%%%%%%%%%%%%%%%%%%%%%%%%%%%%

\section{Concluding remarks}
In summary, using  double-resonance measurements in cold atoms, we have shown that the steady-state populations of  two-level atoms interacting with a phase-modulated microwave field  oscillate at multiples of the modulation frequency. We have applied the Rabi resonance technique to cold atoms, and made the first observations of higher-order harmonics in any system. In the case when the carrier wave is exactly on resonance, we observe that only the even harmonics are present, which is confirmed by numerical solution of modified Maxwell-Bloch equations.
Finally, we find that it is experimentally advantageous to use large optical probe powers in these measurements to damp ``regular'' Rabi oscillations and observe only the response to the modulation.
These observations shed new light on the dynamics of a two-level system, which is currently a workhorse of many practical quantum information applications.

In addition, we have shown that the amplitude of the oscillations at $4\omega_{\rm m}$ has a resonance when $\omega_{\rm m}$ is varied. Our simulations show
that at weak modulation the fourth harmonic has two resonance peaks: when~$\Omega_{\rm R}=2\omega_{\rm m}$~and~$\Omega_{\rm R}=4\omega_{\rm m}$. The quartic dependence of the height of  latter peak and its smaller linewidth compared to that of the Rabi $\beta$-resonance, making a possible alternative for atomic-candle applications in experiments with a weak absorption signal, e.g., in cold atoms.

By exploring in-depth the interactions between atoms and  phase-modulated microwave fields, we have provided the foundations for a high signal-to-noise tool for measuring  microwave fields, and thus the coupling to atomic systems. With the increasing use of 3D microwave cavities~\cite{Reagor2013,Souris2017a}, including with atomic systems~\cite{Adwaith2019}, atomic candle techniques can play an important role in calibrating the microwave field strengths for accurate measures of the coupling strength between the cavity field and, for instance, a microwave qubit~\cite{Grezes2014,Lachance-Quirion2017,Reed2017}. These calibrations will be especially important for techniques that rely on precise timing, such as pulse-area-based quantum memories~\cite{Lvovsky2009,Grezes2016,Saglamyurek2018} and quantum transduction protocols~\cite{Hafezi2012,Mcgee2013,Andrews2015,Kiffner2016}.  

\begin{acknowledgements}
 This work was generously supported by the University of Alberta; the Faculty of Science; the Natural Sciences and Engineering Research Council (NSERC); Alberta Innovates; the Canada Foundation for Innovation (CFI), the Canadian Institute for Advanced Research (CIFAR), and the Canada Research Chairs (CRC) Program. We thank Taras Hrushevskyi, Erhan Saglamyurek and Benjamin Smith for their help with the experimental setup, and thank the Davis lab for the loan of their photodetector.
\end{acknowledgements}

\end{document}